\documentclass[prb,showpacs,preprint]{revtex4-1}
\usepackage{graphicx} 
\usepackage{dcolumn}
\usepackage{bm}
\usepackage{booktabs} 
\usepackage{amsmath}  
\usepackage[colorlinks=true,linkcolor=red]{hyperref} 
\usepackage{memhfixc} 
                       
\begin{document}

\title{Probing Mechanical Properties of Graphene with Raman Spectroscopy}
\author{Nicola Ferralis}
 \email{nferralis@berkeley.edu}
\affiliation{Department of Chemical Engineering, University of California, Berkeley, California 94720}
\date{05/24/2010}

\begin{abstract}

The use of Raman scattering techniques to study the mechanical properties of graphene films is reviewed here. The determination of Gr\"uneisen parameters of suspended graphene sheets under uni- and bi-axial strain is discussed and the values are compared to theoretical predictions. The effects of the graphene-substrate interaction on strain and to the temperature evolution of the graphene Raman spectra are discussed. Finally, the relation between mechanical and thermal properties is presented along with the characterization of thermal properties of graphene with Raman spectroscopy.

\end{abstract}


\maketitle

\section{Introduction}
\label{sec:intro}

The growing interest in understanding the mechanical properties of graphene films is sparked by the ability to control such properties, and thus to modify the structure and electronic behavior for graphene-based applications. Raman spectroscopy is increasingly used to measure accurately and non-destructively graphene mechanical or thermal properties, such as strain or thermal conductivity. This review outlines the current state-of-the-art in the use of Raman spectroscopy to characterize the strain and temperature effects in exfoliated and epitaxial graphene. The relationship between strain and film morphology is also reviewed. 

In section \ref{sec:structure} we review the basic atomic structure of graphene, with a brief overview of the methods used to isolate and prepare graphene films on various substrates. An overview of the mechanical properties of graphene films determined by nanoindentation methods is presented in section \ref{sec:mechprop}, along with the current limitations of such approach. The Raman spectrum of graphene in conjunction with its phonon spectrum is described in section \ref{sec:raman}. A detailed overview of the use of Raman spectroscopy for the determination of mechanical properties of graphene is presented in section \ref{sec:mechpropraman}, with particular emphasis on the characterization of strain and of the temperature effects in the graphene films.

\section{Graphene atomic structure}
\label{sec:structure}
\begin{figure}
\includegraphics{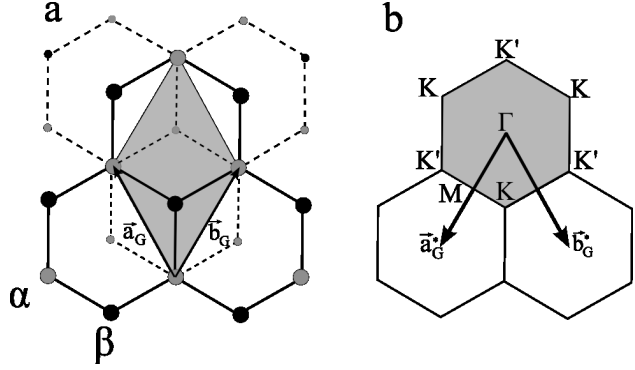}
\caption{\label{Fig1} a) The real space unit cell of a bilayer graphene film in Bernal stacking is shown in gray shade, with basis vectors $\vec{\text{a}}_G$ and $\vec{\text{b}}_G$. Large circles represent the atoms in the top layer, while the smaller ones in the second bottom layer. Atoms in grey and black represent the two inequivalent sublattices $\alpha$ and $\beta$ respectively. b) The reciprocal space unit cell of a single layer graphene, highlighting the high symmetry points and the reciprocal space unit vectors, $\vec{\text{a}}^*_G$ and $\vec{\text{b}}^*_G$. The first Brilloin zone is represented by gray shade.}

\end{figure}

Graphene is a two dimensional sheet of carbon atoms arranged in a honeycomb atomic configuration. A single graphene sheet can be folded and, multiple layers can be  folded or stacked to form sp$^{2}$ carbon in 0D (fullerenes), 1D (carbon nanotubes, CNT) or 3D (graphite). The standard in-plane unit cell of basis vectors $|\vec{\text{a}}_G| = |\vec{\text{b}}_G| = 2.4589\pm0.0005~\text{\AA}$ at 297 K \cite{Baskin_PR1955} contains two carbon atoms (Fig. \ref{Fig1}a). The resulting two dimensional carbon density is 3.820 atom$\cdot\text{\AA}^{-2}$ \cite{Hass_JPCM2008}. Due to the hybridization of carbon bonds into a sp$^{2}$ configuration, each carbon is bonded to three neighboring atoms in a planar configuration. Two sublattices can be identified within a graphene lattice, depending on the orientation of the carbon bonds relative to that of their nearest neighbors (Fig. \ref{Fig1}a). The partially filled $\pi$ orbitals, perpendicular to the graphene plane, are responsible for the electron conduction and the weak interaction between a graphene layer and the underlying substrate. This weak interaction is of the van der Waals type, independent of the substrate\cite{Schabel_PRB1992}. Three possible stacking configurations exist to form graphitic materials, depending on the relative orientation of the graphene layer stacks: Bernal or AB... stacking, hexagonal or AA... stacking, and rhombohedral or ABC... stacking. In this review, we only consider the Bernal stacking for multilayer graphene films, since the mechanical properties of multilayer graphene have been investigated only for this configuration (a comprehensive description of the other stacking sequences can be found in ref \cite{Hass_JPCM2008}). The Bernal (or AB...) configuration is the most common in single crystal graphite (80\%) by virtue of the lowest stacking energy \cite{Haering_CJP1958}. The Bernal configuration is formed by stacking two graphene sheets rotated by 60$^\circ$ relative to each other about the z axis. The three-dimensional unit cell has 4 atoms, and a third basis vector perpendicular to the graphene layer stacks $|\vec{\text{c}}_{G}| = 6.672~\text{\AA}$ at 4.2~K and $6.708~\text{\AA}$ at 297~K \cite{Baskin_PR1955}. The interlayer distance is $\text{c}_{G}/2$. Because of the 60$^\circ$ rotation between the subsequent layers, the two sublattices in graphene see a different local environment in the Bernal configuration: an $\alpha$ atom is positioned directly above an $\alpha$ atom in the sheet below, whereas a $\beta$ atom is positioned above the (empty) center of the ring of the sheet below (Fig. \ref{Fig1}a). The presence of a non-graphitic substrate alters the equivalence between the two sublattices with possible effects on both the mechanical and electronic properties, as discussed in Section \ref{sec:mechpropraman}.

The Brillouin zone for a single graphene layer is shown in Fig.~\ref{Fig1}b. It exhibits high symmetry points: the $\Gamma$ point at the zone center, the M point in the middle of the hexagonal sides and the K and K$^\prime$ points at the corners of the hexagons. K and K$^\prime$ are inequivalent points, since they correspond to the two different and inequivalent sublattices in the graphene atomic structure. 

Graphene samples can be prepared by mechanical exfoliation of highly oriented pyrolithic graphite (HOPG) \cite{Novoselov_S2004, Novoselov_PNAS2005, Geim_NM2007}, which leads to the production of micrometer scale single and multilayer graphene sheets with high degree of control over their thickness. Graphene can be also grown epitaxially on SiC surfaces by high temperature Si sublimation, in ultrahigh vacuum (UHV) \cite{Forbeaux_PRB1998,Berger_JPCB2004} and in controlled environment \cite{Virojanadara_PRB2008,Seyller_NM2009,Jernigan_NL09,Robinson_NL09}. Epitaxial graphene can also be grown on the surfaces of various metals such as Pt \cite{Ueta_SS2004}, Ni \cite{Kim_N2009,Reina_NL2009}, Ir \cite{Coreaux_NL2008,NDiaye_NJP2008}, Ru \cite{Marchini_PRB2007,Parga_PRL2008,Sutter_NM2008} and Cu \cite{Ruoff_S2009}. With this method, large domains can be obtained (domain size $\sim10\mu$m) \cite{Kim_N2009}. Epitaxial graphene grown on metals can be transferred from the synthesis substrate to any chosen substrate \cite{Kim_N2009}. This procedure is suitable for investigation of large scale graphene layers either suspended or transferred to various substrates. The graphene-substrate interaction strongly depends on the type of substrate due to the different degree of adhesion of graphene to the substrate (whether, for example, graphene is grown epitaxially on a substrate or mechanically transferred to it). Therefore the choice of substrate and synthesis method have several implications in the mechanical properties of the epitaxial graphene film.

\section{Graphene mechanical properties measured by nanoindentation}
\label{sec:mechprop}

The mechanical behavior of graphene layers can be described macroscopically by continuum elasticity theory. In this spirit, nanoindentation techniques are well suited to measure the macroscopic mechanical properties of graphene, including Young's modulus and bending stiffness. For example, by using nanoindentation methods on suspended multilayer graphene flakes, the bending stiffness has been measured and found to be in the range from $2 \times 10^{-14}$ N/m to $2 \times 10^{-11}$ N/m for 8 to 100 layers, respectively. Static nanoindentation experiments based on the deflection of AFM cantilevers pressed within 100~nm of the center of $\sim1\mu$m long double-clamped graphene films, provided a measurement of the effective spring constant of multilayer graphene (1-5~N/m). The spring constant was found to scale with the dimensions of the suspended region and the layer thickness (from 5 to 30 layers), and of the extracted Young modulus of 0.5~TPa, independent of thickness \cite{Frank_JVSTB2007}. A significant limitation of the use of nanoindentation techniques is the requirement of a graphene layer to be suspended. The presence of a substrate, over which graphene may either be deposited (SiO$_{2}$ \cite{Novoselov_PNAS2005}, glass and sapphire \cite{Calizo_APL2007b} or polymers \cite{Kim_N2009,Mohiuddin_PRB2009}) or directly grown epitaxially (e.g. SiC \cite{Berger_JPCB2004,Hass_JPCM2008} and metals\cite{Marchini_PRB2007,Sutter_NM2008,Parga_PRL2008}), makes it hard to separate by nanoindentation measurements the intrinsic mechanical properties of a graphene from that of the substrate. 

In contrast to nanoindentation, Raman spectroscopy provides access to information relating to the underlying chemical bonds. Besides complementing the coarse grained approach of macroscopic elasticity, the interrogation of bond vibrations by optical spectroscopy enables the retrieval of information about mechanical and structural properties of films that can have monolayer thickness and be strongly interacting with a substrate.  Raman spectroscopy has thus been used to measure mechanical properties of graphene films, both freestanding and on a substrate \cite{Mohiuddin_PRB2009,Ferralis_PRL2008}, at room and at elevated temperatures  \cite{Calizo_NL2007,Calizo_APL2007}. 

\section{Raman scattering in graphene and graphite}
\label{sec:raman}

\subsection{Raman Spectroscopy of Graphene}

\begin{figure}
\includegraphics{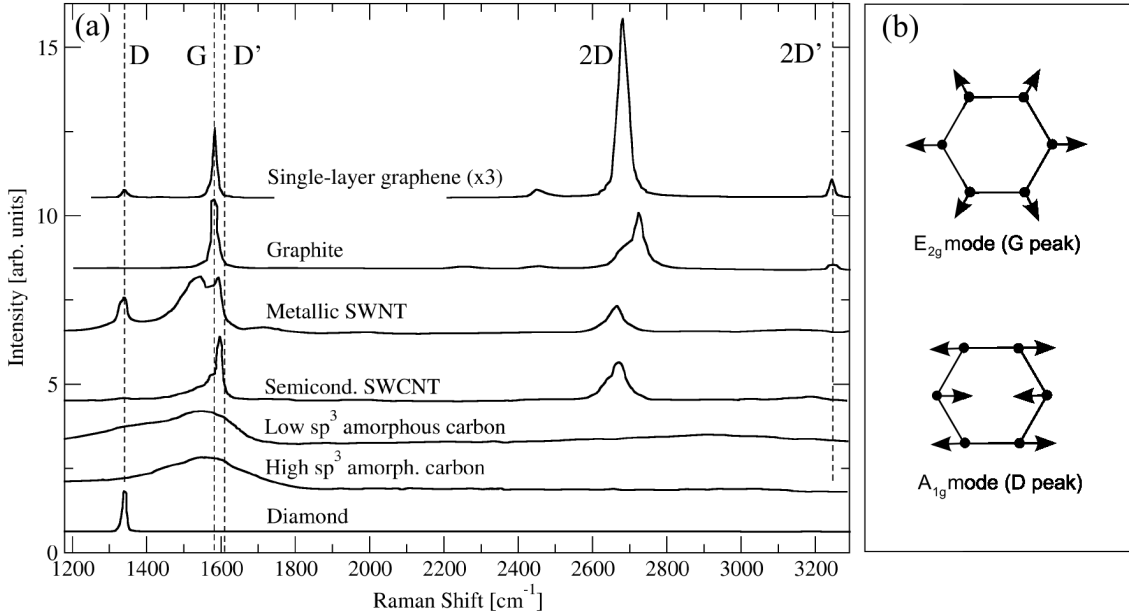}
\caption{\label{CRaman} a) Raman spectra of graphite\cite{Ferrari_SSC2007}, single-layer graphene\cite{Ferrari_PRL2006}, metallic and semiconducting carbon nanotubes\cite{Ferrari_SSC2007}, low and high sp$^{3}$ amorphous carbons\cite{Ferrari_SSC2007}, and diamond\cite{Solin_PRB1970} for visible excitation (excitation energy: 514 cm$^{-1}$). b) "Molecular pictures" of the the E$_{2g}$ and A$_{1g}$ modes, corresponding to the G and D peaks, respectively.} 
\end{figure}  

The Raman spectrum of carbon based materials is characterized by a set of common features in the region between 800 and 2000~cm$^{-1}$, in particular the so-called D and G bands, which lie at around 1330-1360 and 1580 cm$^{-1}$ respectively for visible excitation \cite{Ferrari_SSC2007,Pimenta_PCCP2007,Malard_PR2009}, as shown in Fig.~\ref{CRaman}a. Under these excitation conditions, the Raman spectra of carbon films are dominated by the sp$^{2}$ sites, because visible excitation always resonates with the $\pi$ states. Due to the comparatively small cross-section for the amorphous sp$^{3}$ vs sp$^{2}$ C-C vibrations, a significant fraction of sp$^{3}$ bonds is required in a sample for the sp$^{3}$ peak at 1332 cm$^{-1}$ to be visible, as is the case in diamond (Fig.~\ref{CRaman}) \cite{Ferrari_SSC2007}.

\begin{figure}
\includegraphics{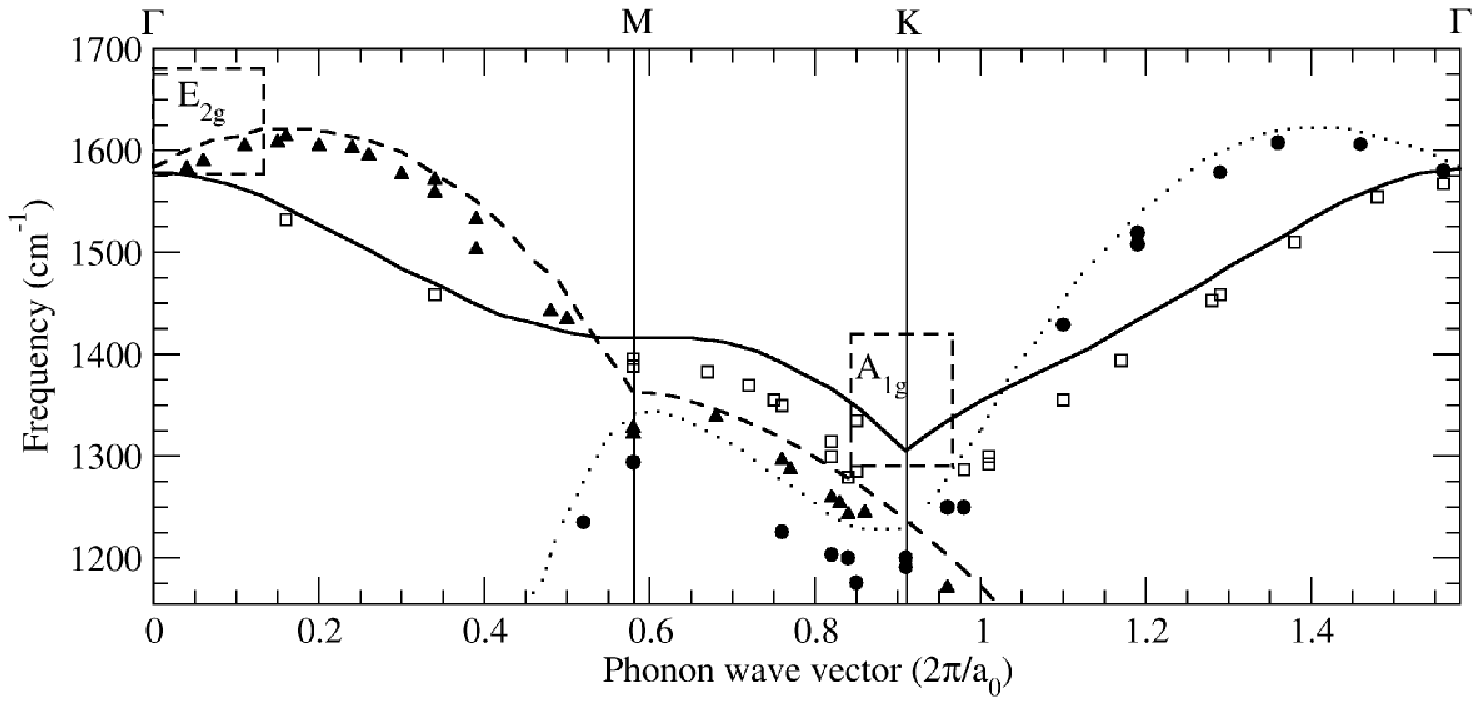}
\caption{\label{GDispCurve} Phonon dispersion plot of a single-layer graphene, calculated (lines) \cite{Piscanec_PRL2004} and experimental (points) \cite{Maultzsch_PRL2004}. Different experimental points corresponds to the different branches.}
\end{figure} 

The phonon dispersion curves of graphene (Fig. \ref{GDispCurve}) are the key to understand its Raman spectrum. They consist of three acoustic phonon modes (A) and three optical (O) phonon modes since the graphene unit cell contains two carbon atoms (Fig.~\ref{Fig1}a). Among these modes, one acoustic branch and one optical phonon branch correspond to out-of-plane phonon modes (o), while for the other acoustic and optical phonon branches, the vibrations, and thus the phonon modes, are in-plane (i). Each in-plane mode has two branches, one longitudinal (L) and one transverse (T). Following the high symmetry $\Gamma$M and $\Gamma$K directions, the six phonon dispersion curves are assigned to LO, iTO, oTO, LA, iTA, and oTA phonon modes \cite{Pimenta_PCCP2007,Malard_PR2009}. In graphite the LO and iTO modes are degenerate at the center of the Brilloin zone, the $\Gamma$ point. According to group theory, these modes are the only Raman active modes, corresponding to the two dimensional E$_{2g}$ phonon. The G peak (located around 1580 cm$^{-1}$) corresponds to such doubly degenerate $E_{2g}$ mode at the Brillouin zone center \cite{Pimenta_PCCP2007,Malard_PR2009}. In the "molecular" picture of carbon materials, the G peak is due to the bond stretching of all pairs of sp$^{2}$ atoms (Fig.~\ref{CRaman}b). 

The D peak ($\sim 1340$~cm$^{-1}$) corresponds to modes associated with transverse optical (iTO) phonons around the edge of the Brillouin zone (\textbf{K} or Dirac point) \cite{Ferrari_SSC2007}. In the molecular picture, it is associated with the breathing mode of the sp$^{2}$ aromatic rings (Fig.~\ref{CRaman}b) \cite{Tuinstra_JCP1970,Ferrari_PRB2000}. The D peak is energy dispersive, so that its position is dependent on the excitation energy (Fig.~\ref{Thickness}) \cite{Pocsik_JNCS1998}. The D peak is usually very intense in amorphous carbon samples, while it is absent in perfect graphitic samples. Its overtone (2D, $\sim 2660-2710$~cm$^{-1}$) however is always visible even when the D peak is absent. Such peculiar behavior is due to the double resonance (DR) activation mechanism \cite{Thomsen_PRL2000} of the D peak, which requires the presence of defects for its initiation \cite{Ferrari_PRB2000,Ferrari_PRB2001}. In a double resonance process, Raman scattering is a four-step process: (i) a laser induced generation of an electron-hole pair; (ii) electron-phonon scattering with an exchanged momentum q$\sim$K; (iii) electron scattering from a defect, whose recoil absorbs the momentum of the electron-hole pair; (iv) electron-hole recombination \cite{Ferrari_SSC2007}. The requirements of conservation of energy and momentum can only be satisfied if a defect is present. In a perfect sample, momentum conservation would be violated by the DR mechanism, and thus the D peak is absent. Momentum conservation however is always satisfied in case of the 2D peak, without the need for defect activation, since the process involves two phonons with opposite momentum vectors\cite{Ferrari_SSC2007}. A similar process is possible with scattering within the same \textbf{K} point. This intra-valley process activates phonons with small momentum q, resulting in the so-called D$^\prime$ peak, located around $\sim 1620$~cm$^{-1}$ in defective graphite \cite{Nemanich_PRB1979}. 

Scattering from holes can also occur in the Raman process. In graphene, under these circumstances, the electron is not scattered back by a phonon of momentum -q, but instead a hole is scattered forward by a phonon with momentum +q. In this case, during the electron-hole generation, both electron and hole scattering processes are resonant. The electron-hole resonant recombination at the opposite side with respect to the K point is also resonant, resulting in the triple resonance scattering process (TR). It has been suggested that the higher intensity of the 2D peak relative to the G band in a graphene monolayer is due to the triple resonance activation mechanism \cite{Malard_PR2009}.

\subsection{Graphene Metrology with Raman Scattering}
\begin{figure}
\includegraphics{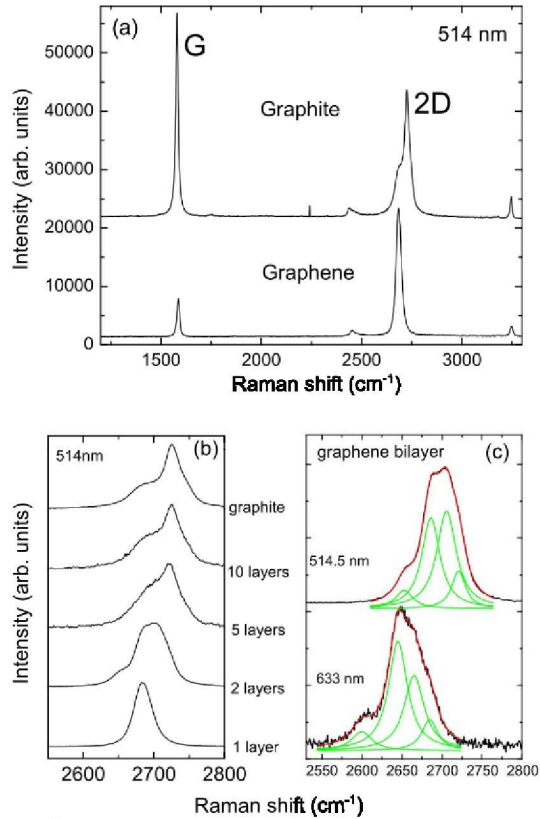}
\caption{\label{Thickness} a) Raman spectra of bulk graphite and single layer exfoliated graphene, taken with excitation energy 514~nm. b) Evolution of the Raman 2D spectra with layer thickness, taken with laser excitation 514~nm. c) The 2D peak in the graphene bilayer is composed of four Lorentzian components, while the single layer has only one. The dispersive nature of the 2D peak is clearly visible in the net shift of the 2D peak in plots c, with excitation $\lambda$ = 514 nm, when compared to its position for an excitation of $\lambda$ = 633 nm \cite{Ferrari_PRL2006}. (Reprinted with permission from ref. \cite{Ferrari_PRL2006}. Copyright 2006 American Physical Society.)}
\end{figure}

Raman spectroscopy, as a non-invasive probing technique, has been extensively employed to characterize graphene layer thickness \cite{Ferrari_PRL2006,Gupta_NL2006}, domain grain size \cite{Tuinstra_JCP1970,Concado_APL2006,Ferrari_SSC2007}, doping levels \cite{Malard_PRB1999,Pisana_NM2007,Ferrari_SSC2007,Das_NN2008,Casiraghi_APL2007}, the structure of graphene layer edges \cite{Concado_PRL2004,Graf_NL2007,You_APL2008,Casiraghi_NL2009}, anharmonic processes and thermal conductivity \cite{Bonini_PRL2007,Balandin_NL2008}.
This has been possible through a combined investigation of the Raman peaks D, G and 2D in graphite and graphene films of various thicknesses and morphologies. An indicative comparison of the Raman spectra of graphene and bulk graphite is made in Fig.~\ref{Thickness}a \cite{Ferrari_PRL2006}. The most striking difference between the individual graphene layers and graphite resides in the change in shape and intensity of the 2D peak. While the 2D peak in graphite consists of two peaks 2D$_{1}$ and 2D$_{2}$ (with intensities of 1:4 and 1:2 compared to the G peak, respectively), the 2D peak in one single graphene layer has only one component with roughly four times the intensity of the G peak (Fig.~\ref{Thickness}a). For multilayer graphene (Fig.~\ref{Thickness}b), the evolution in the shape of the 2D peak has been used to determine the layer thickness \cite{Ferrari_PRL2006,Gupta_NL2006,Graf_NL2007}. The splitting of electronic bands in bilayer graphene is responsible for the splitting of the 2D peak into four components\cite{Ferrari_PRB2000} (Fig.~\ref{Thickness}c). The two lower components further decrease while the higher wavenumber components increase as the film thickness approaches 5 layers. Above this threshold, however, the determination of the layer thickness with Raman becomes rather difficult, as the shape of the 2D peak is increasingly similar to that of bulk graphite. 

Early investigations of disorder in graphitic carbon \cite{Tuinstra_JCP1970} show that the ratio of the D and G band intensities (I$_{D}$/I$_{G}$) is inversely proportional to the in-plane crystallite size L$_{a}$, measured independently with x-ray diffraction. Such relation, known as the Tuinstra-Koenig (TK) relation, has been refined in recent years to provide an empirical method to determine the size of graphene domains from the Raman spectrum under a given excitation energy \cite{Concado_APL2006,Pimenta_PCCP2007}. There are known limitations in this approach, as the distribution of domains with different sizes is such that the smaller domains are weighted more, leading to an underestimation of the average size distribution. In addition, the use of peak intensity ratio instead of peak area ratio, underestimates the average domain size, since the full-width-half-maximum (FWHM) of the D peak increases significantly in comparison to that of the G peak \cite{Ferrari_SSC2007}. Furthermore, the ratio I$_{D}$/I$_{G}$ is known to depend on the electron concentration (and thus on the film doping) \cite{Das_NN2008}, limiting the application of the TK relation when the doping concentration is unknown. Regardless of the limitations, the use of the TK relation allows an estimation of the degree of disorder in the graphene film.

\section{Probing mechanical properties of graphene with Raman spectroscopy}
\label{sec:mechpropraman}
Any changes in the atomic structure in a crystalline solid due to plastic deformation, strain, or thermal expansion are reflected in the phonon spectrum of the crystal. By probing the phonon spectrum with Raman spectroscopy, such changes can be detected, thus providing insight into the mechanical and thermal properties of materials such as graphene. 

Strained semiconductors have received significant interest in the past because of the wide ranging implications of strain, such as the ability to engineer the electronic structure and to affect the carrier mobility in silicon-based materials for electronic device application \cite{Maiti_SSE2004}. The application of an external stress on a crystal results in a lattice strain, i.e., in a change in interatomic distances and consequent redistribution of electronic charge. Isotropic compression (hydrostatic pressure) generally results in an increase in the frequency of the vibrational mode (phonon hardening), while isotropic tension results in the decrease in the vibrational frequency (phonon softening). Application of anisotropic stress has more complex effects, and can result in lifting of the degeneracy of phonon frequencies.
 
In graphene, changes in the Raman spectra have been observed as a consequence of the presence of stress, either induced artificially on suspended or exfoliated graphene\cite{Ni_ACSN2008,Ni_ACSN2008b,Yu_PCCL2008,Ni_PRB2008,Mohiuddin_PRB2009,Chen_NL2009,Metzger_NL2010} or provided by the interaction with the substrate for graphene grown epitaxially on SiC substrates \cite{Ni_PRB2008,Ferralis_PRL2008, Ferralis_APL2008, Rohrl_APL2008,Robinson_NL2009}. Such changes consisted of a systematic upshift in the position of the main Raman D (when present), G, and 2D peaks, by up to 30, 31 and 64 cm$^{-1}$, respectively \cite{Mohiuddin_PRB2009}, for an applied strain of up to 1.3~\%.

\begin{figure*}
\includegraphics{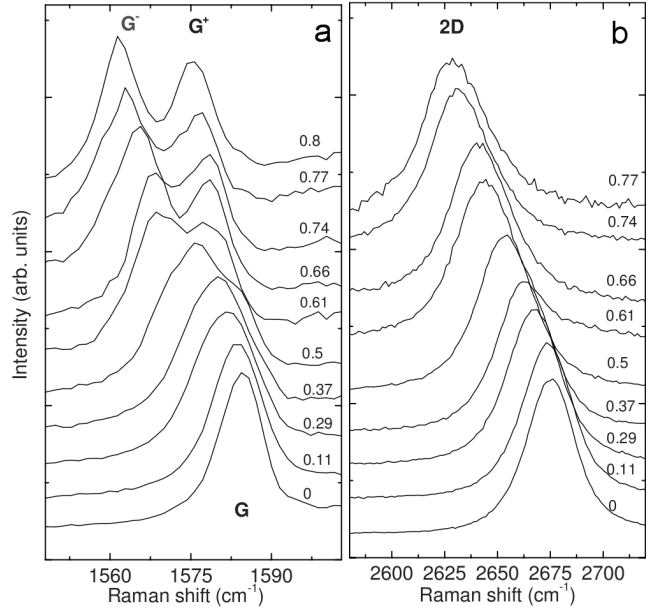}
\caption{\label{StrainDG} Raman spectra in the single-layer graphene of the a) G and b) 2D peaks, as function of the applied uniaxial compressive strain percentage, indicated in the right side of each spectrum. The spectra are acquired with the polarization of the incident light parallel to the direction of the strain. The double degeneracy of the G band is broken as a consequence of the applied strain, resulting in two peaks G$^+$ and G$^-$ \cite{Mohiuddin_PRB2009}. (Reprinted with permission from ref. \cite{Mohiuddin_PRB2009}. Copyright 2009 American Physical Society.)}
\end{figure*}
  
When a uniaxial tensile stress is applied to a graphene layer, the splitting of the G peak has also been observed, reaching up to 15 cm$^{-1}$, for an applied strain of 1.3\% \cite{Huang_PNAS2009,Mohiuddin_PRB2009}. Each peak in the split G band corresponds to two orthogonal modes, having eigenvectors perpendicular to the applied strain (E$^{+}_{2g}$) and parallel to it (E$^{-}_{2g}$). When the uniaxial compressive strain is applied, sp$^2$ bonds along the direction parallel to the applied strain are shortened and hardened, while those perpendicular to it are only slightly affected (Fig. \ref{StrainDG}). Hence, under uniaxial strain, only the peak G$^-$ corresponding to the E$^{-}_{2g}$ mode is significantly shifted relative to the unstrained E$^{0}_{2g}$ (by as much as 30 cm$^{-1}$), while the peak G$^+$ corresponding to the E$^{+}_{2g}$ mode is only moderately shifted (up to 15 cm$^{-1}$). Since this effect is purely mechanical \cite{Mohiuddin_PRB2009}, the full-width-half-maxima of G$^-$ and G$^+$ remain constant. The FWHM of the 2D band is also unchanged. A similar behavior is observed in carbon nanotubes, where the tube curvature induces the splitting of the G band, with a significantly larger shift for the component parallel to the curvature \cite{Piscanec_PRB2007}. The intensities of the two peaks G$^-$ and G$^+$ vary with the polarization of the scattered light along the direction of the strain, allowing the sample crystallographic orientation with respect to the strain to be probed \cite{Huang_PNAS2009,Mohiuddin_PRB2009}. 

In spite of specific changes in the electronic and vibrational band structure, the strain-induced frequency shifts of the Raman active E$_{2g}$ and 2D modes are independent of the direction of strain, which has been observed experimentally \cite{Mohiuddin_PRB2009} and confirmed by \textit{ab initio} calculations \cite{Mohr_PRB2009}.~Thus, the amount of strain can be directly determined from a single Raman measurement \cite{Mohr_PRB2009}.

\subsection{The Gr\"uneisen Parameter for Uni- and Biaxial Strain}
The rate of change with strain of a given phonon frequency in a crystal is determined by its Gr\"uneisen parameter \cite{Grimvall_1986, Mounet_PRB2005}. In metrology applications, accurate values of Gr\"uneisen parameters are crucial for quantifying the amount of strain in the system, reflected in the change in phonon frequency from its value in the absence of strain. In presence of uniaxial strain, the Gr\"uneisen parameter for a particular band $m$ associated with in-plane Raman active phonon band (where $m$ is either the D or G band in graphene), is defined as \cite{Grimvall_1986}:

\begin{equation}
\gamma_m=-\frac{1}{\omega^0_m}\frac{\partial \omega^h_m}{\partial \epsilon_h}
\label{eq1}
\end{equation}
where $\epsilon_h=\epsilon_{ll}+\epsilon_{tt}$ is the hydrostatic component of the applied strain with $ l $ and $ t $ referring to the directions parallel and perpendicular to the applied strain respectively, and $\omega^0_m$ and $\omega^h_m$ correspond to the phonon frequencies of peak $ m $ at zero strain and in presence of an applied strain, respectively. For a given shear component of strain, $\epsilon_s=\epsilon_{ll}-\epsilon_{tt}$, the shear deformation potential $ \beta_m $ is defined as:

 \begin{equation}
\beta_m=\frac{1}{\omega^0_m}\frac{\partial \omega^s_m}{\partial \epsilon_s}
\label{eq2}
\end{equation}

For the G band corresponding to the $E_{2g}$ phonon, the shifts in the two components G$^+$ and G$^-$ relative to the position at zero strain, $\omega^0_G$, are given by:

 \begin{equation}
 \begin{split}
\Delta\omega^\pm_G & =\Delta\omega^h_G\pm\frac{1}{2}\Delta\omega^s_G\\
 & =-\omega^0_G\gamma_G(\epsilon_{ll}+\epsilon_{tt})\pm\frac{\omega^0_G}{2}\beta_G(\epsilon_{ll}-\epsilon_{tt})
 \end{split}
\label{eq3}
\end{equation}
where $\Delta\omega^h_G$ and $\Delta\omega^s_G$ are the shifts associated with the hydrostatic and shear components of the strain respectively. Under condition of uniaxial strain, $\epsilon_{ll}=\epsilon$ and $\epsilon_{tt}=-\nu\epsilon$, where $\nu$ is the Poisson ratio \cite{Grimvall_1986}. In case of graphene, if the layer adheres well to the substrate used for strain analysis, such as for example polyethyleneterephtalate (PET) \cite{Mohiuddin_PRB2009}, the Poisson ratio of the substrate must be used, instead of in-plane Poisson ratio for bulk graphite. Under uniaxial strain, Equations (\ref{eq3}) can be solved, yielding both the Gr\"uneisen parameter and the shear deformation potential for the G band, as functions of the shifts in the positions of the two components G$^+$ and G$^-$:

\begin{equation}
\gamma_G^{uniax}=-\frac{\Delta\omega_G^++\Delta\omega_G^-}{2\omega_G^0(1-\nu)\epsilon}
\label{eq4}
\end{equation}

\begin{equation}
\beta_G=-\frac{\Delta\omega_G^+-\Delta\omega_G^-}{\omega_G^0(1+\nu)\epsilon}
\label{eq5}
\end{equation}
 
Under the conditions of biaxial strain, $ \epsilon_{ll}=\epsilon_{tt}=\epsilon $, there is no shear deformation potential and no splitting of the G peak. In this case, Equation \ref{eq3} can be solved to provide the Gruneisen parameter \cite{Ferralis_PRL2008,Mohiuddin_PRB2009}:

\begin{equation}
\gamma_G^{biax}=-\frac{\Delta\omega_G}{2\omega_G^0\epsilon}
\label{eq6}
\end{equation}

It is however possible that local anisotropies in the applied biaxial strain, possibly induced by the substrate over small domain size (such as in epitaxial graphene grown on SiC), may cause an increase in the FWHM as a result of a local splitting of the G band. It is also worth noting that under biaxial strain conditions, the shift in the peak position is independent of the presence of any substrate, because of the absence of a sheer deformation term and thus the absence of the Poisson term $\nu$ in eq. (\ref{eq6}) \cite{Mohiuddin_PRB2009}. 

The Gr\"uneisen parameter can be similarly derived for the D and D' bands in graphene. Of the two, only the first is single-degenerate, and corresponds to A$ _{1g} $ phonons at the K point (Fig.~\ref{GDispCurve}). The D peak is thus not expected to split under uniaxial strain, and only the hydrostatic component of the stress is present. The Gr\"uneisen parameter for the D peak (which is equivalent to that of the overtone 2D) can be written as:

\begin{equation}
\gamma_{D,2D}^{uniax}=-\frac{\Delta\omega_{D,2D}}{\omega_{D,2D}^0(\epsilon_{ll}+\epsilon_{tt})}
\label{eq7}
\end{equation} 
or:
\begin{equation}
\gamma_{D,2D}^{uniax}=-\frac{\Delta\omega_{D,2D}}{\omega_{D,2D}^0(1-\nu)\epsilon}
\label{eq8}
\end{equation} 

(Note that the shear deformation potential $\beta$ for the D and D' bands cannot be extracted, because of the lack of shear component of the applied uniaxial strain). The D' band is associated with an E symmetry mode, which is double-degenerate; as such, a splitting is expected under uniaxial strain. Experimentally the only study to report on the effects of strain on the D' peak did not observe any splitting, due to the weak intensity of this peak and the small range of applied strain \cite{Mohiuddin_PRB2009}. For small strains, the Gr\"uneisen parameter for the D' follows eq. (\ref{eq8}).
In the case of biaxial strain, equation (\ref{eq7}) is the same as equation (\ref{eq4}), which can be generalized as:

\begin{equation}
\gamma_m^{biax}=-\frac{\Delta\omega_m}{\omega_m^0(\epsilon_{ll}+\epsilon_{tt})}
\label{eq9}
\end{equation}

where $ m $ corresponds to the D, G or 2D bands. It is worth mentioning that in all cases, the detection of strain effects is the most sensitive if the 2D band is considered. With a spectrometer resolution of $\sim2~\text{cm}^{-1}$, the sensitivity for uni- and biaxial strain is 0.03 and $\sim0.01$, respectively.

\begin{table}
\begin{tabular}{|ccc|c|c|c|c|}
	\hline
   $ \gamma_G $  & $ \gamma_D $ & $ \gamma_{D'} $ & $ \beta_G $ & Strain & & Ref.  \\
	 \hline
  1.99  & 3.55 & 1.61 & 0.99 & uniaxial & exp & \cite{Mohiuddin_PRB2009} \\
  2.4 & 3.8 & - & - & biaxial & exp & \cite{Metzger_NL2010}\\
	1.87  & 2.7 & - & 0.92 & uniaxial & th & \cite{Mohiuddin_PRB2009} \\
	1.8  & 2.7 & - & - & biaxial & th & \cite{Mohiuddin_PRB2009,Mounet_PRB2005} \\
	1.72-1.9 &-  & - & - & biaxial & exp (graphite) & \cite{Hanfland_PRB1989,Zhenxian_JPCM1990,Sandler_PRB2003}\\
	 \hline
	- & 2.84$^*$ & 1.74$^*$ & - & biaxial & exp (graphite) & \cite{Goncharov_JETP1990} \\

	\hline
\end{tabular}
\caption{Gr\"uneisen parameter and shear deformation potential for a single layer graphene. The Gr\"uneisen parameter $\gamma_{2D}$ is equivalent to that of $\gamma_{D}$. $^*$ $\gamma_{2D}$ has been measured directly only in case of biaxial strain\cite{Goncharov_JETP1990}.}
\label{tab1}
\end{table}

\subsection{Determination of the Gr\"uneisen Parameter in Graphene}
Mohiuddin \textit{et al.} provided a complete characterization of the Gr\"uneisen parameters for the G and 2D bands of exfoliated graphene\cite{Mohiuddin_PRB2009}. In order to measure the Gr\"uneisen parameters and the shear deformation potential of a single layer exfoliated graphene, equations~(\ref{eq4}), (\ref{eq5}) and (\ref{eq7}) were used to fit measured shifts in the positions of the G and 2D Raman bands as a function of applied uniaxial stress. The resulting Gr\"uneisen parameters and shear deformation potential for the main Raman peaks for a single layer graphene are summarized in table~\ref{tab1}, along with previous theoretical and experimental studies. The values of $ \gamma_G $ and $ \beta_G $ from ref. \cite{Mohiuddin_PRB2009} are in good agreement with those calculated with density-functional theory ($ \sim1.8 $ \cite{Mounet_PRB2005}) and first principle calculations (1.87 \cite{Mohiuddin_PRB2009}). Metzger \textit{et al.} measured directly the Gr\"uneisen parameters for biaxial strain, by placing a single-graphene layer onto a substrated prepatterned with shallow depressions\cite{Metzger_NL2010}. The adhesion of graphene to the substrate across the pre-patterned depression, despite the induced biaxial strain, allowed a controlled and precise determination of the biaxial strain. By using eq.~(\ref{eq9}), the Gr\"uneisen parameters for the G (2.4) and D (3.8) peak were extracted and found to be higher than those measured from uniaxial strain\cite{Mohiuddin_PRB2009} or calculated\cite{Mohiuddin_PRB2009, Mounet_PRB2005}. It has been speculated that the larger values of both the Gr\"uneisen parameters and the peak shifts when compared to previous measurements were due to a better adhesion of graphene to the substrate in the latter studies. This leads to a measurement of the strain actually transferred to the graphene layer from the substrate (i.e., with no slippage). However, the large difference between the measured values of the Gruneisen parameters for the G band (1.8\cite{Mohiuddin_PRB2009} vs. 2.4\cite{Metzger_NL2010}) but not for that of the D peak (3.55\cite{Mohiuddin_PRB2009} vs. 3.8\cite{Metzger_NL2010}) remains unexplained. Recently, Tsoukleri \textit{et al.} investigated the effects of slippage of the graphene layer from the substrate by measuring the strain in a graphene layer, either supported on a poly(methyl methacrylate) (PMMA) cantilever beam or embedded into the cantilever \cite{Tsoukleri_SM2009}. The values of $\partial\omega_{2D}^{uniax}/\partial\epsilon$ were measured by altering the stress applied to the cantilever. The measured values for either the supported or embedded case are in good agreement with those measured initially by Mohiuddin \textit{et al} \cite{Mohiuddin_PRB2009}, suggesting a negligible role of slippage. When compressive stress is applied, however, the effects of buckling and the formation of ridges in the supported graphene layer determine a significantly different value of $\partial\omega_{2D}^{uniax}/\partial\epsilon$, compared to the embedded case. 

By applying such parameters to eq.~(\ref{eq4}), the gradients in Raman peak position per unit of applied strain are extracted. A summary of both theoretical and experimental studies is reported in table~\ref{tab2}. The use of the correct value for the Gr\"uneisen parameter is extremely important, because it affects the estimated value of the Raman peak shift for a given strain. Often the Gr\"uneisen parameter of CNT is used, leading to a questionable estimate for the gradient in the peak position. For example, $ \gamma_{2D}=1.24 $~\cite{Ni_ACSN2008b} $\partial\omega_{2D}^{uniax}/\partial\epsilon \sim -27.1 ~\text{cm}^{-1}/\%$ \cite{Ni_ACSN2008b} to be contrasted to $\sim -83 ~\text{cm}^{-1}/\%$ when $\gamma_{2D}=3.55$ is used per ref.~\cite{Mohiuddin_PRB2009}, obtained on a single-layer graphene. This result has been used to justify the measured value of the gradient in peak position for uniaxial strain. However, the absence of any splitting of the G peak and lack of any difference in Raman peak position between uni- and biaxial graphene \cite{Ni_ACSN2008b}, which are in contradiction with theory, suggests that the applied strain is either far from being uniaxial \cite{Ni_ACSN2008b} or points to poor sample quality. As a further indication, the estimated gradient in the G peak position as function of applied strain ($\partial\omega_G/\partial\epsilon \sim-27.8~\text{cm}^{-1}/\%$) is consistent with the averaged value of the gradients of the shifts in the G$^+$ and G$^-$ peaks ($\partial\omega_G^{uniax}/\partial\epsilon \sim -27~\text{cm}^{-1}/\%$, \cite{Mohiuddin_PRB2009}). This is also consistent with the average value obtained from measurements on carbon fibers ($\sim -25$~cm$^{-1}/\%$), where individual sub-bands cannot be distinguished due to the broad G band for amorphous carbon \cite{Galiotis_JMSL1988}. The similarity in such measurements between graphene and graphite indicates that the in-plane Young modulus for graphene and bulk graphite are similar \cite{Lee_S321}.    

\begin{table}
\begin{tabular}{|cc|cccc|c|c|c|}
	\hline
   $\frac{\partial\omega_G^-}{\partial\epsilon}$ & $\frac{\partial\omega_G^+}{\partial\epsilon}$ & $\frac{\partial\omega_G}{\partial\epsilon}$ & $\frac{\partial\omega_D}{\partial\epsilon}$ & $\frac{\partial\omega_{D'}}{\partial\epsilon}$ & $\frac{\partial\omega_{2D}}{\partial\epsilon}$ & Strain &  & Ref.\\
	 \hline
 -36.4 & -18.6 & - & -41.5 & -22.5 & -83 & uniaxial & exp & \cite{Mohiuddin_PRB2009}   \\
 - & - & -63 & -85.5 & -52 & -191 & biaxial &  exp &\cite{Mohiuddin_PRB2009}   \\
 - & - & -77 & - & - & -203 & biaxial & exp & \cite{Metzger_NL2010}\\
  - 30 & -10.3 & - & -30 & - & -60 & uniaxial & th & \cite{Mohiuddin_PRB2009}   \\
 - & - & -58 & -72 & - & -144 & biaxial & th &\cite{Mohiuddin_PRB2009}   \\
 - & - & - & - & - & -59.1 & uniaxial (supported) & exp &\cite{Tsoukleri_SM2009} \\
 - & - & - & - & - & -65.9 & uniaxial (embedded) & exp &\cite{Tsoukleri_SM2009} \\
 - & - & - & - & - & +25.8 & uniaxial$^*$ (supported) & exp &\cite{Tsoukleri_SM2009} \\
 - & - & - & - & - & +59.1 & uniaxial$^*$ (embedded) & exp &\cite{Tsoukleri_SM2009} \\
 - & - & -14.2 & - & - & -27.1 & uniaxial & exp & \cite{Ni_ACSN2008b} \\
 -34 & -15.4 & - & - & - & -46...54 & uniaxial & th & \cite{Mohr_PRB2009}\\

	\hline
\end{tabular}
\caption{Gradients in Raman peaks position per units of applied strain ($\text{cm}^{-1}/\%$), for a single layer graphene. Whenever the D peak was not present in the measured Raman spectra of the single layer graphene, the gradient in the shift of the D peak is taken as half that of its overtone, the 2D peak, as expected from eq.~(\ref{eq7}) and (\ref{eq9}). $^*$ Measurements carried out under compressive strain. }
\label{tab2}
\end{table}

\subsection{Substrate-Induced Strain on Graphene}

\begin{figure}
\includegraphics{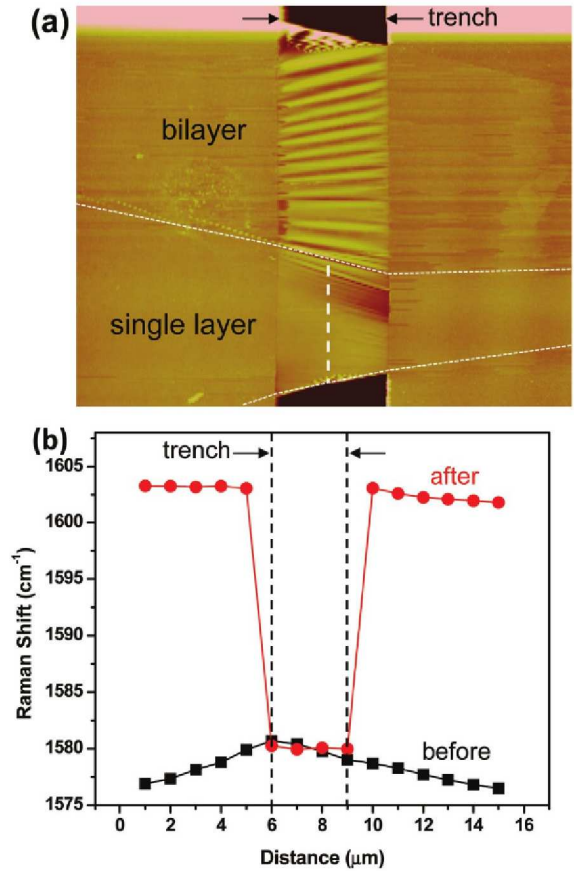}
\caption{\label{Ripples} a) Atomic force micrograph of a graphene layer suspended over a microfabricated trench, after thermal cycling to 700~K. b) Spatial mapping of the G band Raman shift taken perpendicular to the trench before and after the thermal cycling to 700~K \cite{Chen_NL2009}. (Reprinted with permission from ref. \cite{Chen_NL2009}. Copyright 2009 American Chemical Society.)}
\end{figure}

While uni- and bi-axial strain can be artificially applied to suspended graphene layers, strain can arise in graphene heterostructures from the interaction between graphene layers and the underlying substrate. Initially, in the case of exfoliated graphene, no appreciable shifts were observed in the G band of a graphene layer transferred onto SiO$_2$/Si and GaAs substrates \cite{Calizo_APL2007b}. Small downshifts of about 5~cm$^{-1}$ were observed for graphene placed on sapphire and glass, with a splitting of the G band in the latter case \cite{Calizo_APL2007b}. The higher adhesion offered by sapphire substrates, which is sufficient to introduce a small amount of strain in the graphene layer during the mechanical placement, was attributed to the particular binding of carbon-sapphire\cite{Calizo_APL2007b}. The binding is also responsible for the growth of highly aligned CNT on sapphire substrates \cite{Han_JACS2005}. Recently, however, more comprehensive investigations of the evolution of the mechanical and morphological properties in graphene suspended over a microfabricated trench reported variations in the positions of the G and 2D bands, during and after thermal cycling\cite{Chen_NL2009}. Upon thermal cycling to 700~K, while in purely suspended regions no shifts were observed, large upshifts ($\sim23,\sim10,\sim5~$cm$^{-1}$ for a single-, bi- and tri-layer graphene respectively, corresponding to a compressive strain of 0.39\%, 0.18\% and 0.09\% respectively) were instead observed in the regions where graphene was in contact with the underlying substrate. While graphene was compressed in the region over the substrate, the compression was relieved and the formation of ripples was observed in the purely suspended region (Fig.~\ref{Ripples})\cite{Chen_NL2009}.

\begin{figure}
\includegraphics{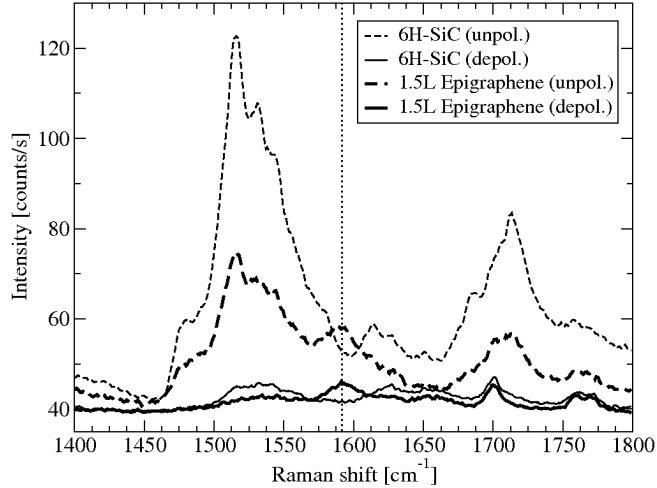}
\caption{\label{GraphSiCG} Raman spectrum of the Si-terminated SiC clean surface is compared with that of 1.5 epitaxial graphene layers. The G peak of graphene (indicated in correspondence to the dotted line, at $\sim$1592~cm$^{-1}$) is convoluted with the second order peaks of the SiC substrate. The scattering contribution of the SiC substrate can be removed by using a depolarized scattering configuration (as shown with the solid lines) \cite{Ferralis_PRL2008}. Excitation energy: 633~nm. (Reprinted with permission from ref. \cite{Ferralis_PRL2008}. Copyright 2008 American Physical Society.)}
\end{figure}

The role of the substrate on strain in graphene films has been also investigated extensively on  graphene grown epitaxially on SiC surfaces (so called epigraphene) by high-temperature decomposition \cite{Ni_PRB2008,Ferralis_PRL2008, Ferralis_APL2008, Rohrl_APL2008,Robinson_NL2009}. Figure~\ref{GraphSiCG} shows representative spectra of a single crystal 6H-SiC(0001) surface and that with 1.5 layers of epigraphene. The Raman peak of zone-center optical (G) phonons in monolayer epigraphene is overwhelmed by the second order signal from the SiC substrate, a broad band occupying the same spectral region. This unfortunate coincidence limits the ability to measure precisely the position of the epigraphene G band itself. This limitation can be overcome by the use of a depolarized scattering configuration \cite{Ferralis_PRL2008}, as shown in Fig.~\ref{GraphSiCG}.

\begin{figure}
\includegraphics{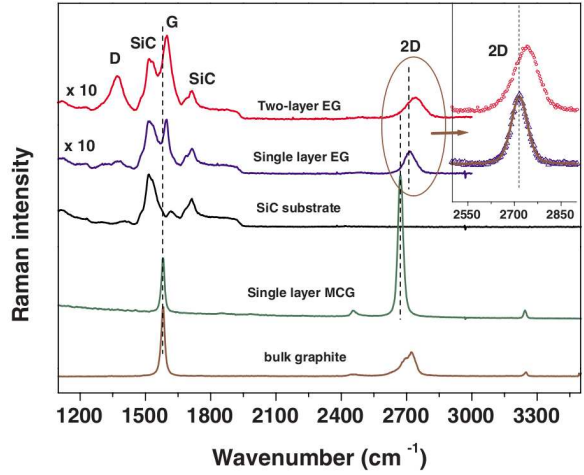}
\caption{\label{RamanEpiG} Raman spectra of single- and bilayer epitaxial graphene on Si-terminated SiC, SiC substrate, micromechanically cleaved/exfoliated graphene (MCG) and bulk graphite as indicated. The shift in the position of the 2D peak is shown in the inset. Excitation energy: 532~nm \cite{Ni_PRB2008}. (Reprinted with permission from ref. \cite{Ni_PRB2008}. Copyright 2008 American Physical Society.)}
\end{figure}

Raman spectra of epigraphene on the Si-terminated 6H- and 4H-SiC (0001) substrates usually show a blueshift in the graphene epilayer peak positions, with respect to those on exfoliated graphene\cite{Ferralis_PRL2008,Rohrl_APL2008}. The extent of shift is different for each Raman peak, as shown in Fig.~\ref{RamanEpiG}. The blueshift recorded varies by up to 22~cm$^{-1}$ for the G peak and 64~cm$^{-1}$ for the 2D peak. Graphene on C-terminated SiC substrates have not been investigated in full details with Raman spectroscopy. However it is speculated that the decoupling of the graphene layer grown on the C-termination may reduce the amount of strain in the film.

The large shift in epitaxial graphene layers on Si-terminated SiC was attributed to compressive strain in the graphene layer. This explanation may seem surprising, since no external strain was applied to the system. However, the only possible alternative explanation, charge transfer from the substrate, was ruled out, based on the fact that it could not account for the magnitude of the shifts in the G and 2D peaks. Indeed, while charging induces a shift in the G peak up to $\sim$20~cm$^{-1}$ for an electron concentration of 4$\times10^{13}$~cm$^{-2}$~\cite{Das_NN2008}, the shift in the G peak corresponding to charge measured in a monolayer graphene on 6H-SiC (1.4$\times10^{13}$~cm$^{-2}$~\cite{Ohta_PRL2007}) would only account for approximately 7~cm$^{-1}$. Similarly, shifts in the 2D band corresponding to the given amount of charge in monolayer graphene is negligible \cite{Das_NN2008}. Hence the observed shifts could only be explained in terms of strain in the system \cite{Ni_PRB2008, Ferralis_PRL2008,Rohrl_APL2008,Mohiuddin_PRB2009}. By using the Gr\"uneisen parameters evaluated under applied uni- and biaxial strain on suspended graphene layers (Table \ref{tab1}), the amount of intrinsic strain in epigraphene can be evaluated using equation (\ref{eq9}). It is interesting to note that the shifts of the D and G peaks occur in the approximate ratio of 1~:~1.4 \cite{Ferralis_PRL2008,Rohrl_APL2008}, which is in good agreement with the ratio between the Gr\"uneisen parameters for those peaks on exfoliated graphene in presence of biaxial stress (1.8~:~2.7, table \ref{tab1}). Hence, for the maximum observed upshift of 22 and 64 cm$^{-1}$ for the G and 2D peaks, the corresponding strain in epigraphene is approximately 0.7-0.8\% \cite{Ferralis_PRL2008}. The shifts in the Raman spectra are found to decrease as the number of graphene layers increases. More specifically, the G and 2D peaks in the epitaxial graphene bilayer are found to be shifted by up to 7 and 22~cm$^{-1}$ (as opposed to 22 and 64~cm$^{-1}$ for the monolayer, respectively), to approach the unstrained values for films thicker than a $\sim$6-9 layers \cite{Rohrl_APL2008}.

\begin{figure}
\includegraphics{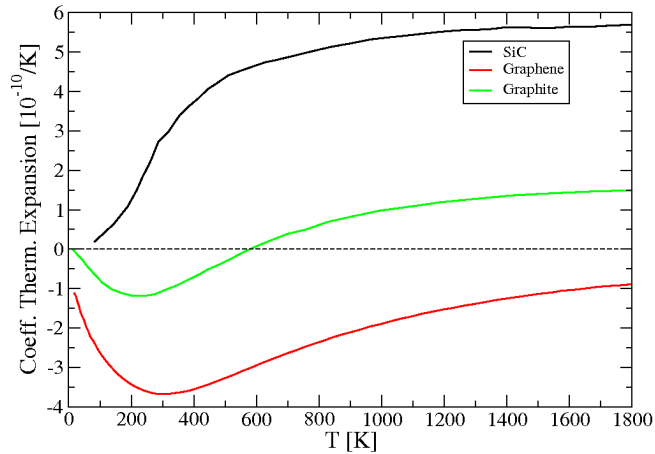}
\caption{\label{CoeffThermExp} Coefficient of thermal expansion as a function of temperature for a single layer of graphene (calculated) \cite{Mounet_PRB2005}, SiC (measured) \cite{Slack_JAP1975}, and graphite. Note that the coefficient of thermal expansion of graphene is always negative.}
\end{figure}

The presence of strain in epigraphene was initially explained in terms of the difference between the lattice constant of the reconstructed 13$\times$13 graphene layer supercell ($\alpha_{graphene}$~=~31.923~\AA) and of the reconstructed SiC $6\sqrt3\times6\sqrt3$ supercell ($\alpha_{SiC}$~=~31.935~\AA) \cite{Ni_PRB77}. Such small difference cannot account for the significant amount of strain measured. Compressive strain at room temperature in the graphene layer was later attributed to the large difference in the coefficients of thermal expansion (Fig.~\ref{CoeffThermExp}) between graphene ($\alpha_{\rm gr}$, as measured and calculated in ref. \cite{Mounet_PRB2005}) and SiC ($\alpha_{\rm SiC}$, as measured in ref. \cite{Slack_JAP1975}) during cooldown from the synthesis temperature \cite{Ferralis_PRL2008,Rohrl_APL2008}. This difference $\Delta\alpha(T)$ is nearly constant between room temperature (RT) and the graphene synthesis temperature, $T_s\approx 1250^{\circ}$C. If the epitaxial film is in mechanical equilibrium with the SiC surface, as a stress-free monolayer commensurate with the 6$\times \sqrt 3$-reconstructed SiC surface at $T_S$, a large compressive strain would develop in the film upon cooling, since SiC contracts on cooling, while graphene expands \cite{Ferralis_PRL2008}:
\begin{equation}
{\frac{1}{1-\epsilon}}=\exp\left[\int_{RT}^{T_{s}}dT' \Delta\alpha(T')\right].
\label{eq_strain}
\end{equation}

\begin{figure}
\includegraphics{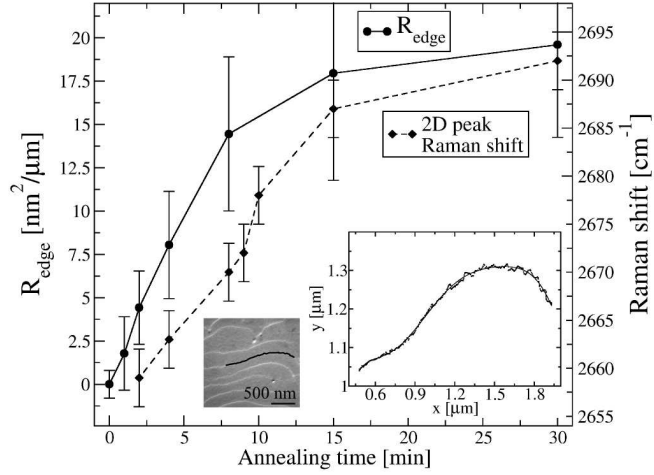}
\caption{\label{RoughTerrace} The evolution of the shift in the 2D peak as a function of the annealing time is compared with the evolution of the edge roughness R$_{edge}$ of the Si-terminated SiC terraces after graphitization \cite{Ferralis_APL2008}. The edge roughness R$_{edge}$ is defined as the difference in the normalized average mean square deviation of any graphitized terrace edge with that of the initial ungraphitized surface. Several profiles of terrace edges are extracted from electron channeling contrast images of samples prepared with the same annealing time and temperature. Each profile (black curve on the ECCI image of a sample annealed for 8 minutes) is fit with a 9$^{th}$ order polynomial to obtain an edge baseline. The normalized average mean square deviation (and thus the edge roughness R$_{edge}$) is extracted from the baseline. (Reprinted with permission from ref. \cite{Ferralis_APL2008}. Copyright 2008 American Institute of Physics.)}
\end{figure}

Ferralis \textit{et al.} found that the shift observed in the position of the Raman peaks strongly depends on the duration of the high temperature annealing \cite{Ferralis_PRL2008,Ferralis_APL2008}. The evolution in the shift of the 2D peak as a function of the annealing time is shown in Fig.~\ref{RoughTerrace}. It was observed that for short annealing times (up to 2 minute) the G and 2D Raman peaks were almost unshifted from their unstrained values. Longer annealing times (up to 1 hour) were found to produce the largest shifts (as high as 22~cm$^{-1}$ for the G band, corresponding to a strain of $\sim$0.8\%, based on eq. \ref{eq9}). It was argued that higher compressive stress at room temperature resulted from a lower stressed film at the synthesis temperature ($T_{S}$), while a nearly stress free film at room temperature indicated that the film existed under high tensile stress at $T_{S}$. Within experimental accuracy, the strain measured at room temperature might well vanish for very short annealing times. In contrast, for long annealing times, the graphene layer reaches mechanical equilibrium with the substrate at the synthesis temperature $T_{S}$, and a compressive strain develops at room temperature film (up to $ \sim $0.8\%). This analysis suggests that mechanical equilibrium with the 6-$\sqrt3$ SiC substrate at $T_{S}$ is indeed achieved for annealing times longer than 10 minutes, while for shorter annealing times ($\sim$5 minutes or less), graphene is under high tensile strain at $T_{S}$ \cite{Ferralis_PRL2008,Ferralis_APL2008}. 

\begin{figure}
\includegraphics{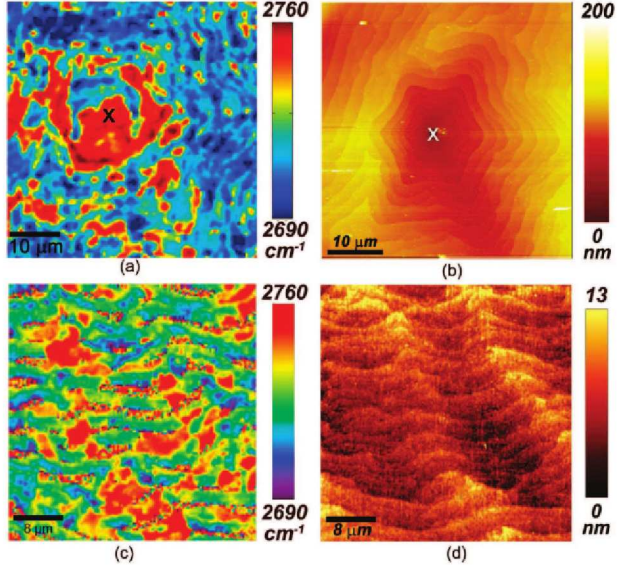}
\caption{\label{MapStrain} Raman spectral map corresponding to the position of the 2D peak of epitaxial graphene grown on the Si-face of SiC. (a) is compared to the AFM image (b) of the same area.  (a, b) near a SiC screw dislocation where its position is marked with an ``x" and (c, d) where such defects are not present. The Raman topography is correlated with the surface morphology of the graphene film as revealed by AFM, suggesting that changes in the physical topography are related to changes in the strain of the graphene film \cite{Robinson_NL2009}. (Reprinted with permission from ref. \cite{Robinson_NL2009}. Copyright 2009 American Chemical Society.)}
\end{figure}

A direct correlation between the strain distribution and graphene surface morphology was made using a combined Raman spectroscopy and electron channeling contrast imaging (ECCI) \cite{Ferralis_APL2008}. It was found that the roughness of the SiC substrate terraces from where epigraphene grows increased paralleling the increase in the Raman peak shifts under the same conditions, as shown in Fig.~\ref{RoughTerrace}. This observation provides a possible mechanism for strain relaxation. For long enough annealing times, tensile strain developed at $T_{S}$ is relieved by the roughening of the step edges to which graphene films are pinned. Such increase in roughness does not induce a significant change in surface coverage ($\pm$0.2~ML). For short annealing times, surface relaxation and roughening do not take place, leaving the SiC terraces morphologically unchanged. Similarly, large inhomogeneities in the distribution of strain within the same epigraphene layer were reported by combined Raman mapping and atomic force microscopy (AFM) \cite{Robinson_NL2009}. Large shifts in the 2D Raman band (up to 74~cm$^{-1}$, corresponding to a strain of about 1.0\%) were observed to correspond to regions with screw dislocations, step terraces and macrodefects, while regions with less pronounced band shifts corresponded to large flat terraces (Fig. \ref{MapStrain}). The strain distribution map obtained with Raman spectroscopy appears to be correlated with the surface morphology of the graphene film, monitored by AFM, confirming  that changes in the physical topography are related to changes in the strain of the graphene film \cite{Robinson_NL2009}.

 \subsection{Characterization of Thermal Properties of Graphene with Raman Spectroscopy}

\begin{figure}
\includegraphics{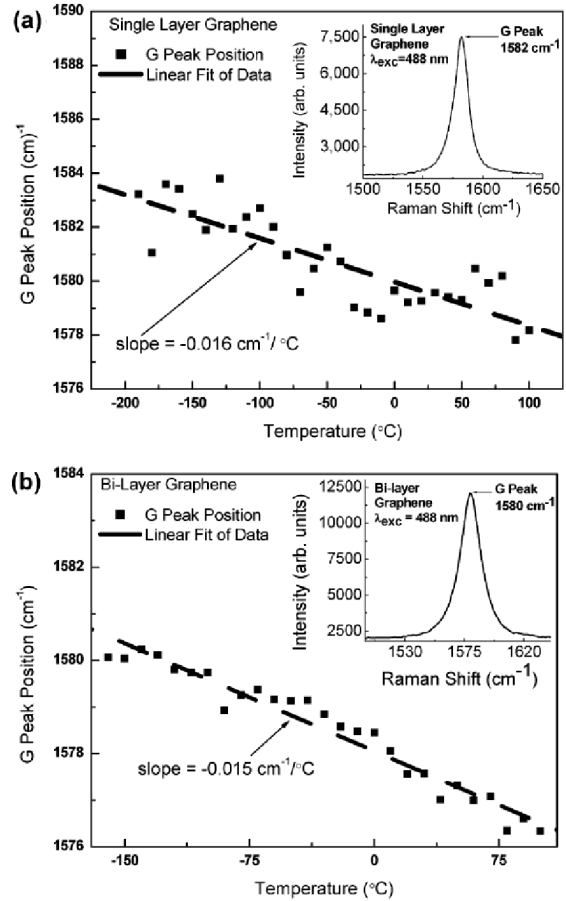}
\caption{\label{TDep} Temperature dependence of the G peak position (shown in the inset) for the single- (a) and bilayer (b) exfoliated graphene. The measured data is used to extract the temperature coefficient for G peak \cite{Calizo_NL2007,Calizo_APL2007}. (Reprinted with permission from ref.\cite{Calizo_NL2007}. Copyright 2007 American Chemical Society.)}
\end{figure}

Raman spectra have a significant temperature dependence, both in intensity and in position of the Raman peaks. For example, the ratio of the intensities of antiStokes and Stokes peaks is commonly used as a metrology tool to determine the actual temperature of the analyzed sample \cite{Gu_JRS96}. Since strain in the lattice also affects Raman peak positions, it is crucial to understand and discern the role played by the changes in lattice parameters (due to strain or thermal expansion) from purely isovolumetric thermal dependencies. Experimentally, separating the two contributions is complicated, especially if either mechanism is not easily controllable, or strictly depends on the position of the Raman peak for its determination. 
In complete absence of strain, shifts in Raman peaks observed in response to temperature changes reflect both elementary anharmonic processes (electron-phonon and phonon-phonon scattering) and changes in lattice parameters with temperature (thermal expansion). The temperature dependence of the G and 2D peak positions $ \omega_m $ for single and bilayer suspended graphene is approximately described by \cite{Calizo_NL2007,Calizo_APL2007}: 

\begin{equation}
\omega_m=\omega^0_m+\chi_m T
\label{eq10}
\end{equation}
where $\omega^0_m$ is the position of the peak $ m $ (either G or 2D) at T=0~K, and $\chi_m$ is the first-order temperature coefficient of the same peak. By measuring the position of the G and 2D peaks as a function of sample temperature, the temperature coefficients are extracted for single and bilayer graphene (Fig. \ref{TDep}). The results are reported in table \ref{tab3}, and compared with other carbon-based materials. It should be noted that the geometrical configuration employed in these experiments (a graphene sheet rigidly connected to the substrate) does not guarantee the conditions of a strain-free environment. Hence, the actual determination of the thermal evolution of the Raman spectrum through these experiments may include non-negligible contribution from strain.

\begin{table}
\begin{tabular}{|c|c|c|c|c|c|}
	\hline
    & Peak & $\chi$  & $\omega^0_m$ & Temperature & Ref. \\
    & & [cm$^{-1}$/K] & & range [K] &\\
	 \hline
   Single layer suspended & G & -0.0162 & 1584 & 83-373 & \cite{Calizo_NL2007}\\ 
   Single layer suspended & \raisebox{-2ex}{G} & \raisebox{-2ex}{-0.040} & \raisebox{-2ex}{-} & \raisebox{-2ex}{400-500} & \raisebox{-2ex}{\cite{Cai_NL2010}} \\
   \raisebox{1ex}{and supported on Au/SiO$_2$} & & & & & \\ 
   Bilayer & G & -0.0154 & 1582 & 113-373 & \cite{Calizo_NL2007}\\[2ex]
   \hline
   HOPG & G & -0.011 & 1584 & 83-373 & \cite{Calizo_APL2007}\\
   SWCNT & G & -0.0189 & - & 299-773 & \cite{Raravikar_PRB2002}\\
   DWCNT & G & -0.022 & - & 180-320 & \cite{Bassil_APL2002} \\
   diamond & G & -0.012  & - & 300-1900 & \cite{Zouboulis_PRB1991}\\  
   \hline
   Single layer & 2D & -0.034 & 2687 & 83-373 & \cite{Calizo_APL2007}\\
   Bilayer & 2D & -0.066 & 2687 & 113-373 & \cite{Calizo_APL2007}\\ 

\hline
\end{tabular}
\caption{Temperature coefficients for the G and 2D peaks in suspended graphene layers. The values of $\omega^0_m$ are extrapolated by fitting \cite{Calizo_APL2007,Cai_NL2010}. The $\chi$ values for the G peak are compared to those for other carbon-based materials.}
\label{tab3}
\end{table}

The temperature dependence of the G peak for the single layer is found to be higher than for the bilayer. Both values are higher than that for HOPG, and are expected to approach the HOPG value for thicker graphene films. The temperature coefficient $\chi_m$ depends on the anharmonic potential constants, the phonon occupation number and the thermal expansion of the graphene two-dimensional lattice \cite{Postmus_PR1968}. The contribution of anharmonic terms is most significant at high temperatures; hence the overall thermal dependency is not expected to follow a linear trend \cite{Bonini_PRL2007}. The non-linearity must be taken into account when using calibration of thermally induced shifts in the Raman spectra of graphene. Commonly used linear fits need to be accompanied by the temperature range used for the measurements, as reported in Table~\ref{tab3}. In HOPG, $\chi_m$ is found to depend mostly on the anharmonic contribution, due to direct coupling of phonon modes. Since thermal expansion occurs primarily along the c-axis, its effect on the in-plane G and 2D Raman modes are not very pronounced\cite{Tan_APL1999}. 

It is however important to note that the interaction with the substrate may strongly affect thermal expansion of graphene, resulting in a different value of $\chi_m$ for purely suspended versus strongly interacting graphene layers. This might be the cause of the different value measured by Cai \textit{et al.} for a single layer graphene grown by CVD and pressed against a Au/SiO$_2$ thin film on Si \cite{Cai_NL2010}. As a further indication of a strong interaction with the substrate, the same value of $\chi$ was found on regions of the same graphene layer either supported and suspended over circular microfabricated holes.    

\begin{figure}
\includegraphics{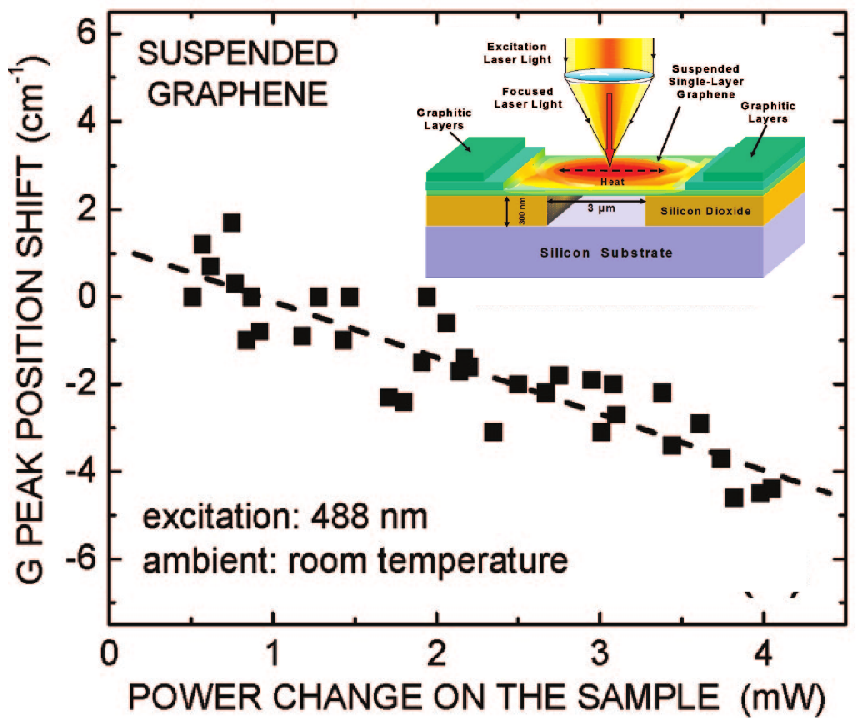}
\caption{\label{TC} The thermal conductivity measurement is performed by monitoring the change in position of the G peak as a function of the total dissipated power. The excitation laser light focused on a single layer graphene suspended across a trench (inset), is to create a local radiative hot spot, and to generate a heat wave across the graphene layer \cite{Balandin_NL2008}. (Reprinted with permission from ref.\cite{Balandin_NL2008}. Copyright 2008 American Chemical Society.)}
\end{figure}

Raman measurements on suspended nanostructures can be used to determine their thermal conductivity. This method has been employed to measure the thermal conductivity of single layer graphene \cite{Balandin_NL2008,Cai_NL2010,Faugeras_ACSN2010} (Fig.~\ref{TC}). In one experiment, a single layer of graphene is mechanically placed across microfabricated SiO$_2$ trenches, to remove any interaction of the graphene layer with the substrate. The laser source used for the Raman measurement is also used as the local heating probe. By monitoring the shift in the G peak as a function of the change in laser power P, the thermal conductivity $ K $ of a graphene layer can be obtained according to\cite{Balandin_NL2008}:

\begin{equation}
K=\frac{\chi_G(L/2hW)}{(\partial\omega/\partial P)}
\label{eq11}
\end{equation}
where $\chi_G$ is the temperature coefficient of the G peak, $L$ is the the distance from the middle of the suspended graphene layer to the heat sink, $h$ and $W$ are the thickness and width of the graphene layer, respectively. Equation~\ref{eq11} is valid under the assumption that the front wave is non spherical, as is usually the case when the laser spot size ($\sim 0.5-1.0~\mu$m) is of the same order as the graphene strip lateral size\cite{Balandin_NL2008}. Although the interaction with the substrate is minimized across the trenches, residual strain may still be present in the supporting regions. The amount of strain in the suspended region however was considered negligible, as the Raman peak position in this region, at room temperature, corresponds to that of unstrained suspended graphene (Fig.~\ref{Ripples})\cite{Chen_NL2009, Bonini_PRL2007}. Furthermore, the coefficient $\chi_G$ in eq.~\ref{eq11} is measured on an unsuspended graphene layer, while the experiment is performed on a suspended layer. $\chi_G$ for an unsuspended graphene monolayer is expected to be lower than that for the suspended layer because of the interaction between the graphene layer and the substrate. Therefore, the measured thermal conductivity is underestimated. As previously noted, Cai \textit{et al.} performed a similar experiment where graphene synthesized via chemical vapor deposition was pressed against a Au/SiO$_2$ thin film on Si \cite{Cai_NL2010}. The significant difference in the value of $\chi_G$ measured in these experiments may be due to an enhanced interaction of graphene with the substrate, possibly due to the graphene synthesis and deposition method. Under these conditions, the coefficient of thermal expansion of graphene is strongly affected by that of the substrate, leading to a value of $\chi_G$ which is significantly different from that of a purely suspended graphene film. Further investigations are needed to quantify how the thermal evolution of the graphene Raman spectra is affected by the graphene-substrate interaction and in particular by the difference in the coefficients of thermal expansions of graphene and the substrate. 

The measured thermal conductivity is compared to those of other carbon based materials in table \ref{tab4}. The extremely high value of phonon thermal conductivity in the strictly two-dimensional graphene layer is in sharp contrast with the reduced phonon thermal conductivity (as compared to bulk values) in quasi-one-dimensional systems such as nanowires \cite{Zou_JAP2001}, or quasi-two-dimensional semiconducting thin films \cite{Balandin_PRB1998}.~The net reduction in the phonon thermal conductivity observed in these systems is explained in terms of rough boundary scattering or phonon spatial confinement effects. Given the high values measured for a single layer graphene, such effects appear not to be present.~Furthermore, when comparing the thermal conductivity of a single layer graphene to other graphitic materials such as CNT, graphene exhibits a higher value, possibly due to a reduced number of structural defects, and a reduced intralayer scattering. In a comparison with bulk graphite, thermal conductivity approaches that of bulk as the number of atomic planes in graphene films increases from 2 to 4 \cite{Ghosh_NM2010}. It has been shown that Umklapp-limited thermal conductivity of graphene grows with the increasing linear dimensions of graphene flakes and can exceed that of the basal planes of bulk graphite when the flake size is on the order of a few micrometers \cite{Nika_APL09,Nika_PRB09}. 

\begin{table}
\begin{tabular}{|c|c|c|c|c|c|}
	\hline
    & & K & Method  & T & Ref. \\
    & & [W/mK] & & [K] & \\
	 \hline
   Single layer & suspended & $\sim$4840-5300 & optical & 300K & \cite{Balandin_NL2008}\\
   Single layer (CVD) & suspended & $\sim$2500 & optical & 350K & \cite{Cai_NL2010}\\ 
   & suspended & 1400 & optical & 500K & \cite{Cai_NL2010}\\ 
   & supported & 370 & optical & 300K & \cite{Cai_NL2010}\\ 
   Single layer & supported & $\sim$600 & electrical & 660K & \cite{Seol_S2010}\\ 
   \hline
   2 layers & suspended & $\sim$2800 & optical & 300K & \cite{Ghosh_NM2010}\\
   3 layers & suspended & $\sim$2250 & optical & 300K & \cite{Ghosh_NM2010}\\
   4 layers & suspended & $\sim$1270 & optical & 300K & \cite{Ghosh_NM2010}\\
   8 layers & suspended & $\sim$1240 & optical & 300K & \cite{Ghosh_NM2010}\\
   \hline
   SW-CNT & - & $\sim$3500 & electrical & - & \cite{Pop_NL2006}\\
   MW-CNT & - & $>$3000 & electrical & - & \cite{Kim_PRL2001}\\
   Diamond & - & 1000-2200 & electrical & - & \cite{Shamsa_JAP08} \\
   Diamond-like carbon & - & $\sim$0.2 & electrical & - & \cite{Shamsa_APL2006} \\
 
\hline
\end{tabular}
\caption{The thermal conductivity of a single and multilayer layer graphene is measured optically via Raman spectroscopy. Thermal conductivities of single layer graphene, single- and multi-wall CNT and diamond (via the 3-$\omega$ method) are showed for comparison.}
\label{tab4}
\end{table}  

\section{Conclusions and Outlook}
\label{sec:conclusions}

Raman spectroscopy is currently used as a metrology tool to determine the extent, the quality and the uniformity of graphene films. This review has illustrated the applications of Raman spectroscopy to probing the mechanical properties of graphene films. The direct measurement of Raman peak shifts, for example, has enabled the determination of parameters such as the Gr\"uneisen parameter and the shear deformation potential, and thus to a measurement of the strain in graphene films. While such shifts, in general, can be attributed to other causes (e.g. induced charge, doping), under precise experimental conditions (thermal equilibrium, constant pressure, and with fixed Fermi level) lattice strain can be directly measured from peak changes in the Raman spectra \cite{Mohr_PRB2009}. Understanding the evolution of strain in graphene films is important, as it allows for a deeper understanding of how graphene interacts with the environment, and particularly with a substrate. The ability to monitor and control strain in graphene could be crucial during device fabrication, as it affects the electronic properties of the material itself  \cite{Pereira_PRL09}. For example, it has been recently shown that modulation in electrical \cite{Teague_NL2009} and optical \cite{Choi_PRB10} conductance can be induced by strain. It has been suggested that by properly modulating strain locally in graphene may lead to a controlled tuning of the electronic band gap \cite{Guinea_NP2009}. Such studies are in their infancy, however. The vast majority of investigations have been performed either on exfoliated graphene, or on epitaxial graphene grown on SiC. More investigations are needed to understand the presence and the evolution of strain in graphene grown, for example on transition metals via chemical vapor deposition, or as an effect of the mechanical transfer in the case of exfoliated graphene. Since deposition or synthesis methods strongly affect the graphene interaction with the substrate, further studies are needed to highlight and establish a connection between the strength of this interaction and the thermal evolution of the Raman spectra of graphene. While attempts to correlate strain to other structural properties of graphene (such as surface morphology) have been proposed, more work is needed to be able to connect strain with the electrical, optical and thermal properties of the material. As doping strongly affects strain in thin films \cite{Zhang_JMM06}, more investigations are required to determine how doping affects the strain in graphene films.  
 
From a fundamental standpoint, Raman spectroscopy can provide accurate \textit{in situ} measurements of thermal properties such as the thermal conductivity. Such approach allows for the characterization of the role of geometry, chemistry, and morphology, and of their effects on thermal properties. Such capabilities need to be extended to other graphene-related materials, such as graphene oxide \cite{Stankovich_N2006,Dikin_N2007} and graphane \cite{Elias_S2009}. When applied to graphene in a controlled environment, these measurements, may prove suitable for sensing applications.  
Overall, the characterization of mechanical properties of graphene with Raman spectroscopy will promise to be valuable in the determination of the optimal growth conditions, and even more in the optimization of fabrication methods of graphene-based devices.

\begin{acknowledgments}
The author is grateful to Roya Maboudian and Carlo Carraro for useful suggestions and their critical reading of this review. This work was supported by the National Science Foundation under Grants CMMI-0825531 and EEC-0832819 through the Center of Integrated Nanomechanical Systems.

\end{acknowledgments}

\bibliographystyle{apsrev4-1} 

%

\end{document}